\documentstyle{article}
\input epsf.sty
\begin{document}
\twocolumn

\section*{Solar Flares: Gamma Rays}

\noindent In press: Encyclopedia for Astronomy (Macmillan)

\subsection*{OVERVIEW} Electromagnetic radiation extends over a very 
broad range of wavelengths, from gamma rays at the shortest 
wavelengths to radio waves at the long-wavelength end of the 
spectrum. In terms of photon energies, gamma rays are at the high 
end of the spectrum, starting at a few tenths of an MeV. This unit 
of energy (1 MeV=1.6$\times$10$^{-13}$joule) is used throughout this 
article. In solar flares, as at many other astrophysical sites, 
gamma-ray emission results from interactions of fast particles with 
an underlying ambient medium. These fast particles, protons, 
$\alpha$ particles, heavier nuclei and electrons, are accelerated 
from the ambient plasma by the electric fields associated with the 
complex and varying magnetic fields in the flaring solar atmosphere. 
Thus, solar flare gamma rays can teach us about the mechanisms that 
accelerate the particles, in particular those which yield particles 
with energies in excess of about 1 MeV, the minimum energy needed to 
produce gamma rays. This is quite important for the understanding of 
flare mechanisms, because such protons and $\alpha$ particles, along 
with lower energy electrons, contain the bulk of the energy released 
in flares. In addition, the solar flare gamma-ray emission exhibits 
characteristic spectral lines which provide information on the 
elemental composition of the ambient solar atmosphere. 

\subsection*{Gamma-ray production mechanisms}

Solar flare gamma-ray emission exhibits both lines and continuum 
(Fig.~1). This theoretical spectrum extends over the entire energy 
range in which gamma rays from flares were observed. The lines 
appear at energies from about 0.5 to 8 MeV, whereas the continuum 
extends up to at least 1000 MeV. Up to about 1 MeV, and again from 
about 10 to 50 MeV, the continuum is dominated by bremsstrahlung 
produced by the braking of the accelerated electrons in the Coulomb 
fields of the ambient nuclei and electrons. The bremsstrahlung 
produced by ultrarelativistic electrons is strongly collimated along 
the direction of motion of the electrons. The lines result from the 
deexcitation of nuclei, from the capture of neutrons, and from the 
annihilation of positrons. The relevant nuclear cross sections are 
available from accelerator measurements. Deexcitation lines are 
either narrow or broad. Narrow lines result from the bombardment of 
ambient nuclei by accelerated protons and $\alpha$ particles, while 
broad lines result from the inverse reactions in which accelerated C 
and heavier nuclei collide with ambient H and He. The strongest 
narrow deexcitation lines are at 6.129 MeV from $^{16}$O, 4.438 MeV 
from $^{12}$C, 1.779 MeV from $^{28}$Si, 1.634 MeV from $^{20}$Ne, 
1.369 MeV from $^{24}$Mg and 0.847 MeV from $^{56}$Fe. The broad 
lines merge into a quasi-continuum above the bremsstrahlung between 
about 1 and 8 MeV. The broadening of the deexcitation lines is the 
consequence of the Doppler shifting of the essentially monochromatic 
radiation produced in the rest frame of the excited nuclei. In the 
case of the narrow lines the broadening is due to the recoil 
velocity of the excited nuclei which is quite small. The widths of 
the broad lines are much larger because the excited nuclei continue 
to move rapidly after their excitation.

\begin{figure}
\begin{center}
\leavevmode\epsfxsize=19pc\epsfbox{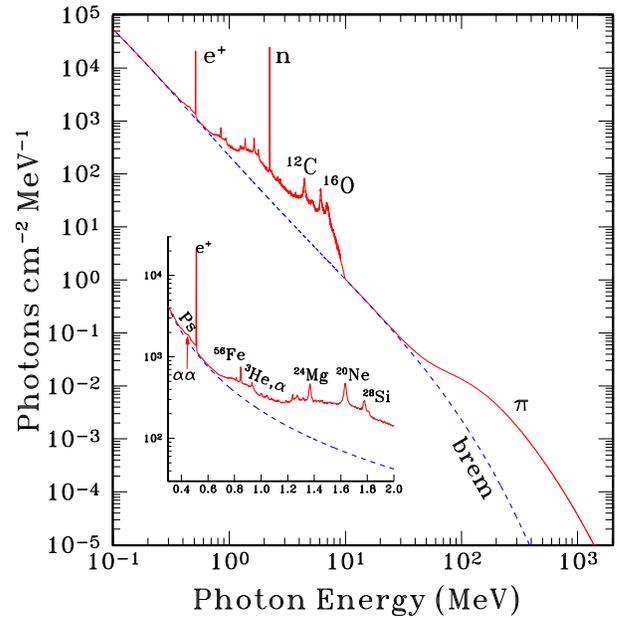}
\end{center}
\caption{Theoretical solar flare gamma-ray spectrum constructed to 
fit the observations.}
\end{figure}

The strong, very narrow line at 2.223 MeV is due to neutron capture. 
All accelerated ions (protons and heavier nuclei) produce neutrons. 
The dominant neutron production mode in solar flares is the breakup 
of He nuclei, both in the accelerated particles and the ambient 
medium. Along with the deexcitation lines, the neutrons are produced 
at sites most likely located in the chromospheric portions of 
magnetic loops. The neutrons propagate both upward, away from the 
Sun, and downward into the photosphere where they are first 
thermalized by elastic collisions with protons and subsequently 
captured mostly by protons to produce deuterium and essentially 
monoenergetic photons at 2.223 MeV, the binding energy of deuterium. 
The neutrons moving away from the Sun can reach Earth where they 
were detected with ground-based and Earth-orbiting instruments. The 
protons resulting from neutron decay in the interplanetary medium 
were also detected.

The 2.223 MeV neutron capture line is very narrow because it is 
broadened by only the relatively low photospheric temperature of 
about 6000 K. Because the production site of the 2.223 MeV line is 
situated much deeper than that of the nuclear deexcitation lines, 
the 2.223 MeV line can be attenuated resulting in limb darkening. 
This means that for flares located at or near the solar limb, the 
intensity of the line is much weaker than that of the deexcitation 
lines, in contrast with disk flares located far from the limb for 
which the 2.223 MeV line is the strongest. A competing mode of 
neutron capture in the photosphere is that on $^3$He. This has been 
used to obtain information on the photospheric $^3$He abundance.

\begin{figure}
\begin{center}
\leavevmode\epsfxsize=19pc\epsfbox{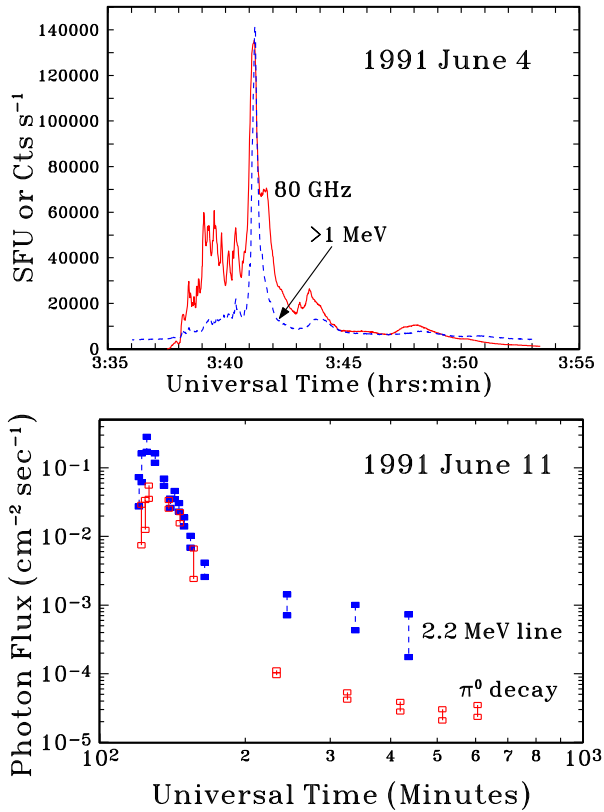}
\end{center}
\caption{Time dependencies. The 
upper panel exhibits the impulsive nature of the gamma-ray emission, 
along with the associated high frequency radio emission. In 
contrast, the lower shows gamma-ray emission extending over many 
hours.}
\end{figure}

Another strong narrow line is that at 0.511 MeV from positron 
annihilation (Fig.~1). The positrons result mainly from the decay of 
various short lived radioactive nuclei, for example $^{11}$C, 
$^{13}$N and $^{15}$O, which are also produced by interactions of 
the accelerated ions. The positrons subsequently either 
annihilate directly into 0.511 MeV gamma rays, or form positronium 
(an atom analogous to hydrogen with the nuclear proton replaced by a 
positron), which also annihilates into gamma rays. Positronium 
annihilation leads to both line emission at 0.511 MeV and continuum 
below this energy. The positronium continuum (denoted by Ps) can be 
seen in Fig.~1 at energies just below the e$^+$ (0.511 MeV) line. 
The width of this line is very sensitive to the temperature of the 
medium in which the positrons annihilate. For the calculations of 
Fig.~1 it was assumed that the positrons annihilate in the 
chromosphere.

Two strong lines result from the interactions of $\alpha$ particles 
with He. Fast and ambient $^4$He nuclei fuse into $^7$Li and $^7$Be 
which are born either in their ground states or in their respective 
excited states at 0.429 and 0.478 MeV. Because of Doppler 
broadening, the ensuing deexcitations produce a relatively broad 
emission feature centered around 0.45 MeV (except under conditions 
of strong accelerated particle anisotropy when the two lines are 
narrowed into separate distinguishable features). The combined 
feature, generally referred to as the $\alpha$$\alpha$ line, can be 
seen in Fig.~1, superposed on the Ps and bremsstrahlung continua. 
Along with the $\alpha$$\alpha$ line, there are several other lines 
which can only be excited by accelerated $\alpha$ particles, as well 
as lines which are excited exclusively by accelerated $^3$He nuclei. 
The latter are of interest  because of the very large $^3$He 
abundances observed in accelerated particles from impulsive flares 
(see below). The lines in question are at 1.00, 1.05 and 1.19 MeV, 
and at 0.937, 1.04 and 1.08 MeV, from $\alpha$ particle and $^3$He 
induced reactions, respectively.

At high energies the continuum in some flares is dominated by pion 
decay radiation. Neutral and charged pions are produced mostly in 
high energy (greater than hundreds of MeV) proton-proton, 
proton-$\alpha$ particle and $\alpha$-$\alpha$ interactions. Neutral 
pions decay directly into two photons, while charged pions decay 
(via muons) into secondary electrons and positrons which produce 
gamma rays via bremsstrahlung and annihilation in flight. The 
combined pion decay radiation is shown in Fig.~1. 

\subsection*{The data and their implications}

Gamma-ray lines from solar flares were first observed in 1972 with a 
detector flown on spacecraft. But it was not until 1980 that routine 
observations of gamma-ray lines and continuum became possible with 
the much more sensitive spectrometer on the Solar Maximum Mission 
(SMM), a spacecraft that carried out successful solar observations 
for almost a decade. During that period, gamma-ray observations were 
also carried out with a smaller instrument on the Japanese 
spacecraft HINOTORI. During the 1990's, solar flare gamma rays have 
been detected with instruments on the COMPTON GAMMA RAY OBSERVATORY 
(CGRO). This observatory was launched in 1991, and it is expected 
that it will continue to operate well into the first decade of the 
21$^{\rm st}$ century. Additional solar gamma-ray observations 
during this period were carried out with instruments on the GRANAT 
and GAMMA-1 spacecraft, which are no longer operational, as well as 
with a small detector on the YOHKOH spacecraft which continues 
functioning. Starting in 2000, a new spacecraft, the High Energy 
Solar Spectroscopic Imager (HESSI), will carry out solar flare X-ray 
and gamma-ray observations. The main implications of the already 
available data are the following: 

\subsubsection*{Flare energy release} 

Hard X-ray observations of solar flares demonstrated that a major 
fraction of the released flare energy resides in sub-relativistic 
electrons of energies above 0.02 MeV. This, together with the 
observed impulsiveness of the hard X-rays, strongly suggested that 
electron acceleration to these subrelativistic energies is closely 
associated with the process that releases the flare energy initially 
stored in magnetic fields. But prior to the availability of 
gamma-ray data, the accepted paradigm was that ion acceleration is 
only a secondary manifestation of the flare energy release process. 
The gamma-ray emission, however, turned out also to be very 
impulsive. Moreover, recent studies based on the relative 
intensities of gamma-ray lines (in particular the $^{20}$Ne line at 
1.634 MeV) provided new information on the energy distribution of 
the accelerated ions, requiring very large particle fluxes near 1 
MeV. It was shown that the energy contained in such ions is 
comparable to that contained in the subrelativistic electrons. It 
thus appears that a large fraction of the released flare energy 
(approximately 10$^{32}$ ergs for large flares) is indeed in 
accelerated particles, but equipartitioned between ions and 
electrons. The top panel in Fig.~2 shows the impulsive time profile 
of gamma-ray emission produced by ions and electrons of MeV energies 
compared with very high frequency radio emission produced by 
electrons of similar energies gyrating in solar magnetic fields of 
hundreds of gauss.

\subsubsection*{Particle acceleration and transport at the Sun}

Particles accelerated at or near the Sun are also observed by 
detectors on spacecraft in interplanetary space. These observations 
have led to the identification of two classes of acceleration 
events, impulsive and gradual. Among the various characteristics of 
the two classes, the composition of the accelerated particles is 
perhaps the most important. The impulsive events exhibit large 
enhancements of relativistic electrons relative to MeV protons, of 
$^3$He relative to $^4$He, and of heavy ions (particularly Fe) 
relative to the C and O. In contrast, the gradual events have 
smaller electron-to-proton ratios ($e/p$), and their heavy ion 
abundances and $^3$He-to-$^4$He ratios are similar to coronal 
values. The strong association of the gradual events with coronal 
mass ejections (CME) suggest that the particles in these events are 
accelerated by CME driven shocks. On the other hand, the electron, 
$^3$He and heavy ion enrichments in impulsive events require 
selective acceleration which is most likely achieved by gyroresonant 
interactions with plasma waves.

The gamma-ray observations have independently revealed the 
characteristics of impulsive acceleration. In particular, high $e/p$ 
ratios are required by the observed continuum to line flux ratios, 
high $^3$He abundances are suggested both by the 2.223 MeV line 
observations, which require enhanced neutron production, and very 
recent findings in SMM data of $^3$He induced lines. In addition, 
there is evidence in the GRANAT data for highly enhanced heavy ion 
abundances based on broad lines from Ne, Mg, Si and Fe. 

The increased sensitivity of the CGRO detectors, and the occurrence 
of large flares while these instruments were observing the Sun, have 
shown that flares can produce gamma rays for very long periods of 
time. The data, showing gamma ray emission lasting for up to 8 hours 
after the impulsive phase of the flare, can be seen in the bottom 
panel of Fig.~2. It is still not known whether the particles were 
accelerated in the impulsive phase of the flare and subsequently 
trapped in magnetic loops at the Sun or accelerated continuously 
over the duration of the emission. 

Additional information on accelerated particle transport at the Sun 
was obtained from observations of the 2.223 MeV line and electron 
bremsstrahlung above 10 MeV. The limb darkening of the 2.223 MeV was 
observed from many solar flares, and it was demonstrated most 
dramatically by gamma-ray observations of a flare located 10$^\circ$ 
behind the limb for which the 2.223 MeV line was absent 
while the deexcitation lines were still seen. Evidently, a 
considerable fraction of the interactions occurred in the corona, at 
a site which was visible from earth orbit while the neutron capture 
site in the photosphere was occulted. On the other hand, there is 
one observation of a behind-the-limb flare from which the 2.223 MeV 
line was seen. Because of the very strong expected attenuation, the 
observed 2.223 MeV line must have been produced by charged particles 
interacting on the visible hemisphere of the Sun. These particles 
were either accelerated by a coronal shock over a large volume, 
thereby producing an extended gamma-ray emitting region, or 
accelerated locally at the flare site whence they propagated along 
large loops to the visible hemisphere.  

In contrast to the limb darkening of the 2.223 MeV line emission, 
the bremsstrahlung above 10 MeV was observed to be limb brightened. 
This means that the flares from which such emission was observed 
were preferentially located close to the solar limb. This effect is 
most likely the consequence of particle motion in magnetic loops 
which converge toward the footpoints with the particles radiating 
most efficiently when they move parallel to the photosphere near the 
mirror points.

\subsubsection*{Ambient medium abundances}

SMM and CGRO data on narrow gamma-ray lines have provided 
information on solar atmospheric elemental abundances. While the 
C-to-O abundance ratio was found to be consistent with both 
photospheric and coronal values, the Mg-to-O, Si-to-O and Fe-to-O 
ratios turned out to be enhanced relative to the photospheric 
abundances but consistent with those of the corona. The first 
ionization potentials (FIP) of Mg, Si and Fe are lower than those of 
C and O. The enhancement of the abundances of low FIP elements in 
the corona relative to the photosphere has been known from both 
atomic spectroscopy and particle observations of gradual events, but 
the origin of this fractionation is still only poorly understood. 
The gamma-ray results, and the fact that the gamma-ray lines are 
most likely produced in the chromosphere, indicate that the FIP bias 
sets in already at relatively low heights in the solar atmosphere. 
Ongoing research on the $\alpha$$\alpha$ line indicates that in the 
gamma-ray production region either the $\alpha$ particle or the 
ambient He abundances is enhanced, exceeding the standard He/H value 
of 0.1. \vskip 0.2 truecm

The early development of the field is summarized in the reviews of 
Chupp, and Ramaty and Murphy. Much of the recent observations and 
theory, the relationship of the gamma-ray studies to other solar 
flare investigations, as well as a detailed historical review (by 
Chupp), are given in the HESP conference proceeding edited by 
Ramaty, Mandzhavidze and Hua. The Murphy et al. paper provides 
details on ongoing research.

\subsection*{Bibliography}

\noindent Chupp E L 1984 High-Energy Neutral Radiations from the Sun 
{\it Ann. Rev. Astron. Astrophys.} {\bf 22} 359-387 \vskip 0.2truecm

\noindent Murphy R J et al. 1997 Accelerated Particle Composition 
and Energetics and Ambient Abundances from Gamma-Ray Spectroscopy of 
the 1991 June 4 Solar Flare {\it Astrophys. J.} {\bf 490} 883-900 
\vskip 0.2truecm

\noindent Ramaty R, Mandzhavidze N and Hua X M (eds.) 1996 {\it High 
Energy Solar Physics} (New York: AIP) \vskip 0.2truecm

\noindent Ramaty R and Murphy R J 1987 Nuclear processes and 
accelerated particles in solar flares {\it Space Science Rev.} {\bf 
45} 213-268 \vskip 0.2truecm

\subsection*{Authors}

Reuven Ramaty and Natalie Mandzhavidze\par

\end{document}